\documentclass[11pt,fleqn]{article}
\usepackage{latexsym,amsmath,amssymb}
\usepackage{longtable}
\usepackage{caption}
\usepackage{a4wide}
\usepackage{graphics}
\usepackage{float}
\usepackage{latexsym}
\usepackage{color}
\usepackage{amsmath,amssymb}
\usepackage{mathrsfs}
\usepackage{rotating}
\usepackage{amsfonts}
\usepackage{amsmath}
\usepackage{amssymb}
\usepackage{graphicx}
\usepackage{amsmath}
\usepackage{breqn}
\usepackage{epstopdf}
\usepackage{dsfont}
\usepackage{geometry}
\usepackage{multirow}
\usepackage[authoryear]{natbib}
\begin{document}
\numberwithin{equation}{section}
\renewcommand{\thefootnote}{*}
\renewcommand{\figurename}{\bf Fig.}
\thispagestyle{empty}

\begin{center}
{\Large \Large\bf Sentiment Analysis of ESG disclosures
on Stock Market}\\
\vspace{0.5cm}
{\bf Sudeep R. Bapat$^{1,*}$, Saumya Kothari$^1$ and Rushil Bansal$^1$}\\
\vspace{0.2cm}
$^1$Indian Institute of Management Indore, India\\
\vspace{.2cm}
$^*$Email: sudeepb@iimidr.ac.in
\end{center}
\begin{abstract}
\noindent
In this paper, we look at the impact of Environment, Social and Governance related news articles and social media data on the stock market performance. We pick four stocks of companies which are widely known in their domain to understand the complete effect of ESG as the newly opted investment style remains restricted to only the stocks with widespread information. We summarize live data of both twitter tweets and newspaper articles and create a sentiment index using a dictionary technique based on online information for the month of July, 2022. We look at the stock price data for all the four companies and calculate the percentage change in each of them. We also compare the overall sentiment of the company to its percentage change over a specific historical period.
\end{abstract}
{\bf Keywords:} ESG, Sentiment Analysis, Composite Score, Tesla, HSBC Bank, Amazon, Goldman Sachs, Twitter

\section{Introduction}
Impact Investing as an investment strategy has grown tremendously over the past decade, and since then, both the investors as well as significant organizations have focused on the Environment, Social, and Governance aspects of their activities. In this investment style, investors focus on the positive impact created by their investments and the company’s commitment to serving society as a whole, starting from their employees, environment, and ethical practices. The rise of impact investing diversified a company’s focus from just their shareholders to the modern standard of corporate responsibility as well.

In the stock market, a company's share price is said to depend on many factors, one of which is Company Related News. This factor has a wide range of sources, such as newspaper articles, press releases, social media updates like Twitter tweets, etc. We have observed that the effect of the news on such forums has a significant impact on the stock market. Hence, with the advent of impact investing, the focus on such forums has increased as all company activities are critically analyzed and presented to the public with a focus on ESG considerations. This has led to the emergence of various scoring and rating solutions by statistical organizations that have subsequently caused investors to analyze and assess the ESG performance of various companies and integrate the same into their stock market portfolio. However, such parameters for scoring and rating suffer from some limitations. The primary limitation is that these scores are based on the company’s voluntary disclosures of ESG or CSR activities. This leads to a bias in the market as a company tries to conceal information that has a negative connotation. Secondly, the analysis of such activities is done on a qualitative basis which relies on the judgment and opinions of the analyst. Therefore, the reliability of these scores is affected by such limitations and thereby missing the ESG market timeline relevant to the price change.

Hence, we aim to overcome these challenges and present a relationship between ESG Related Disclosures and the real-time stock prices of the company. To facilitate this, we aim to provide real-time Twitter data focusing on the tag “ESG Investing.” Even though the credibility and robustness of Twitter data is established, we reiterate four primary advantages of our Twitter dataset: First, a fat API pipeline receives real-time tweets from Twitter, making our information
source timely. Tweets are forward-looking because they capture expectation formation and hence help us overcome the challenge of time lag. Third, unlike rankings computed on arbitrary guidelines and/or companies' own reports, Twitter is an open speech forum in that it provides everyone who feels entitled a voice on a topic of interest can present their analysis. Fourth, from an investment perspective, most ESG disputes and severe ESG events, such as corporate frauds, have originated from a whistleblower inside or outside the organization. Nowadays, ESG issues are voiced and flagged on social media, especially Twitter, which has become the global "virtual speaker's corner." Hence, Twitter detects anomalies or disruptions in a real-time format. In addition, we have taken newspaper articles as the other sources to diversify the forums and integrate an official source in the study, which also considers the press releases and company updates.

Hence, with the help of the dataset of tweets and news, we aim to calculate the sentiment of each single data point. Subsequently, we will analyze the stock trends of specific companies and form a correlation between the sentiment of the dataset with the price changes we observe in the stock.

\section{Further Relevant Literature}\label{sec2}

Firms' performance in the area of Environmental, Social, and Governance (ESG) issues appear to be particularly actively scrutinized by their investors and the public, as seen by the establishment of sustainable investment funds or demand for ``green finance" \citep{gilbert2019rising}. The amount and quality of data available to explore whether and how investors respond to information about a company's ESG performance has also evolved. As the empirical evidence to answer this question is still scarce and contradictory in the academic literature, this paper seeks to fill this research gap by extracting ESG information from publicly available news articles (Twitter and newspaper article sources) and investigating its relationship with the stock market performance of four companies' stocks namely, Amazon, Tesla, HSBC Bank and Goldman Sachs.

We are basing our research on the work of various authors, all of whom have disputed hypotheses, conclusions, and opinions on the subject. According to \citep{friedman2007social}, a firm's only social responsibility is to generate lawful profits. This viewpoint would imply that any ESG activity that is not part of a company's core business should not be undertaken by the company and that investors should not incorporate ESG-related information into their investment decisions other than by withdrawing capital from companies that engage in such activities. \citep{brammer2006corporate} present supporting evidence for this effect of ESG activities, finding that organizations with greater social performance scores had lower returns than those with lower social performance scores. Similarly, \citep{kruger2015corporate} and \citep{capelle2019every} find that positive ESG-related information can hurt a firm's market value in the near run. \citep{cheong2017reactive} discover that most organizations have a reactionary attitude toward ESG matters, which provides greater insight into the likely mechanism behind such observations. These corporations participate in ESG activities excessively only after experiencing poor market and investor sentiment in the preceding year, where market and investor sentiment is recorded by a modified version of Baker and Wurgler's index \citep{baker2006investor}. While this conclusion would simply have ramifications for how altruistic a company's motivations behind its ESG initiatives are on its own, \citep{goss2011impact} demonstrate the potential consequences of such behavior. The authors suggest that corporations' ESG operations, which are launched in direct response to negative media and investor sentiment, are frequently regarded as window-dressing, lowering the company's perceived creditworthiness and increasing its cost of financing.

Other empirical research, however, provides evidence for some (in)direct financial gains for corporations from being proactive in ESG problems, casting doubt on a wholly gloomy perspective of a firm's ESG operations. \citep{lins2017social}, for example, show that ESG initiatives can increase stakeholder trust, which can subsequently be depended upon in times of economic distress, such as during the 2008-2009 financial crisis. \citep{nofsinger2014socially} provide a similar case, expanding on the financial performance of socially responsible investment funds under various market scenarios. Similarly, \citep{cahan2015corporate} discovered that corporations with a high level of social responsibility have a better overall press image. According to the authors, a positive media image helps to develop a better reputation, boosts investor trust, and may allow the company to realize economic benefits from increasing positive public awareness.

From this brief summary of the ESG literature, it is clear that the link between ESG information and stock market reaction, and consequently investors' reactions to a firm's ESG initiatives, has yet to be firmly established. Assuming that a company's only responsibility is to generate legal profits, ESG initiatives should be regarded as superfluous. As a result, investors who hold this viewpoint should penalize any additional ESG activity information. Even if investors recognize certain favorable externalities of ESG-related actions, they will sell more of their shares. When there is unfavorable ESG news on the market, or when freshly initiated ESG activities are regarded as merely a technique to greenwash a firm's public image. Positive ESG-related news, on the other hand, could be viewed as an intangible asset that increases investor trust and improves a company's reputation or public image. Thus, there appears to be an incentive for a company to position itself as a sustainable, socially responsible firm if investors respect companies' ESG efforts and see value in such activities even if they are not part of the core business. Given the current surge in popularity of the topic and the financial resources allocated to the domain of ESG, with ESG-focused assets under management growing by about 20\% per year, it looks to be critical - for both parties - companies and investors alike - to better understand how investors digest ESG-related information \citep{bank2018big}.

Other studies, for the most part, just skim the surface of the importance of sentiment in determining the impact of ESG activities on business performance (e.g., \citep{cheong2017reactive}, \citep{goss2011impact}, \citep{naughton2014csr}). Linking the information included in a newspaper article on a specific ESG issue, as measured by a sentiment index, with investors' reactions to this information, on the other hand, could be a critical component in furthering our understanding of the relationship between the two. By emphasizing ESG sentiment, we build on the findings of various writers in the behavioral finance literature who have found a strong association between sentiment and stock price fluctuations (e.g., \citep{tetlock2007giving}, \citep{garcia2013sentiment}, \citep{li2014news}). Given the findings of this strand of literature, the question arises as to whether changes in ESG-specific sentiment, i.e., sentiment expressed toward a company's ESG-related actions, have an impact on a stock's financial performance.

\section{Methodology}\label{sec3}
Now let’s understand the model; we have taken the dependent variable as the price fluctuation or changes in the stock price of specific companies and the independent variable as the sentiment of individual newspaper articles and tweets. We have chosen 4 companies to focus on for our dataset. These companies are HSBC Bank, Amazon, Tesla, and Goldman Sachs. We have taken these companies to have a diverse view of the relationship. Each of these companies belongs to a significant sector in the economy. HSBC Bank belongs to the Banking Industry, Amazon to the E-Commerce Industry, Tesla to the Automobiles Industry, and Goldman Sachs to the Finance industry.

The model is based on real-time Twitter data using an API Pipeline and newspaper articles for the past 10 days. We analyze the stock prices for the past 20 days by taking into account the opening price of each day and the volume of stocks traded. The Twitter dataset contains the Timestamp, Tweet ID, User ID, Name, Followers, Place, and most importantly, the Ticker for each specific tweet. We have tried to consolidate the tweets belonging to a specific Ticker, such as for HSBC, it will belong to “ESG Investing HSBC,” and so on. Therefore, with the help of the titles mentioned above, we can guarantee the credibility of the dataset, and the Ticker provides the relevance we require in the data. Similarly, in the newspaper articles, the dataset contains the Timestamp, URL, Title, and Ticker. We have used the same Ticker in both datasets to maintain consistency.

Subsequently, we gathered stock data using real-time data from Yahoo Finance. The stock prices have been streamlined to find relevant data for the past 20 days using function tail(20). With the help of the 20 days data, we will be better able to analyze the effect of Twitter tweets and newspaper articles on the stock as we can see the net price changes for the first 10 days without the sentiment analysis and then observe the changes after taking into consideration the sentiment from each tweet and article. To garner the sentiment of every datapoint, we compiled our datasets of news and tweets and used a pre-trained model in python to determine the sentiment and sentiment score for each tweet and article. Subsequently, for analysis purposes, we assigned the sentiments a score of $1, 0,$ and $-1$ for positive, neutral, and negative sentiments, respectively. Furthermore, we calculate the composite score for each datapoint by multiplying the sentiment value $(1,0,-1)$ and the sentiment score, which ranges from 0 to 1. Once we organized our data of composite sentiment scores of each datapoint and the net stock price changes for the four companies, we moved forward with our findings and conclusion.

\section{Data}\label{sec4}
\subsection{Choice of companies}

For the subsequent analysis, we have taken four companies to focus on for our research. These companies are one of the most visible brands in each of their sectors, such as Banking - HSBC Bank, Automobiles - Tesla, Finance - Goldman Sachs, and E-commerce - Amazon. We have taken the most visible brands for our research as our primary objective is to analyze the effect of ESG on stock prices. ESG in itself is a relatively less explored domain, and therefore, the most visible brands will provide the most accurate data as the information is available publicly and resonates with the people quickly.

\subsection{Dataset of newspaper articles and twitter tweets}
We have taken a combined data set of newspaper articles and tweets as with the growing trend of social media, public influence would be garnered quickly with Twitter data, and for daily updates and ESG disclosures, we have taken the newspaper articles datasets. We have gathered live data in the selected time frame using an API pipeline which helped us extract any relevant tweet or newspaper article according to the Ticker established before and fell under the domain of Environment, Social, or Governance (ESG).

\subsection{Graphical representation of stock prices}
The four companies - HSBC Bank, Tesla, Amazon, and Goldman Sachs - have been carefully segregated, and their prices for the twenty days have been extracted using Yahoo! Finance, various important metrics such as their opening price, closing price, the stock volume of trade, date, and time have been extracted. However, for our study, we have only examined the opening price, date, and volume.

Each graph in Figure 1 depicts the stock prices in the form of a candlestick graph, showing the highs and lows of the stock every day for the selected time period. From the stock trends for every company, we can observe that there is a rising trend for all four stock prices in the given time period.

\[%
\begin{tabular*}
{\textwidth}[c]{@{\extracolsep{\fill}}cc}%
$\includegraphics[width = .45\textwidth]{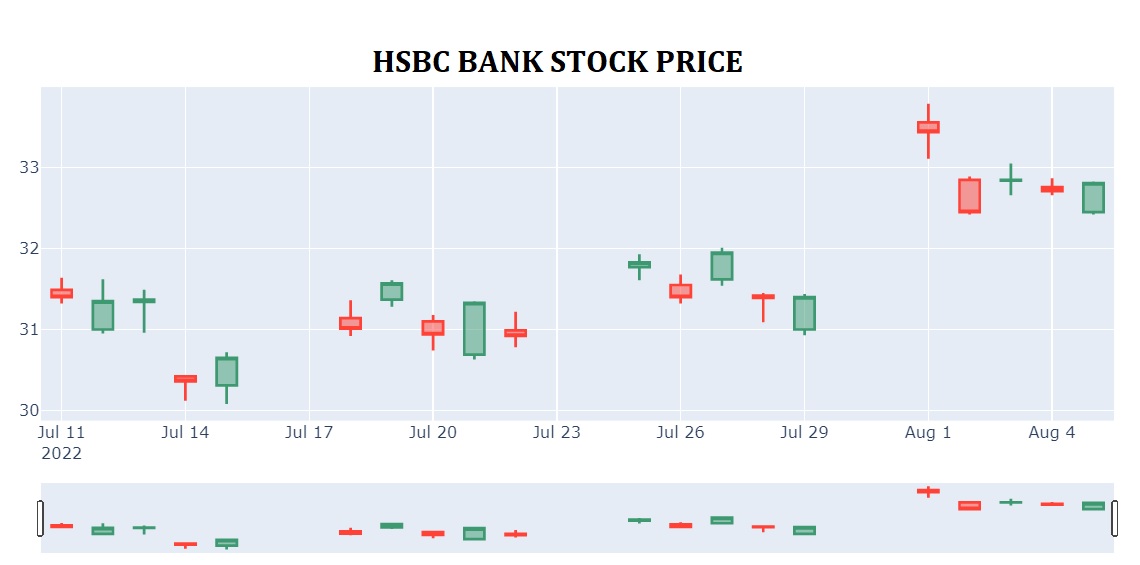}$ & $\includegraphics[width = .45\textwidth]{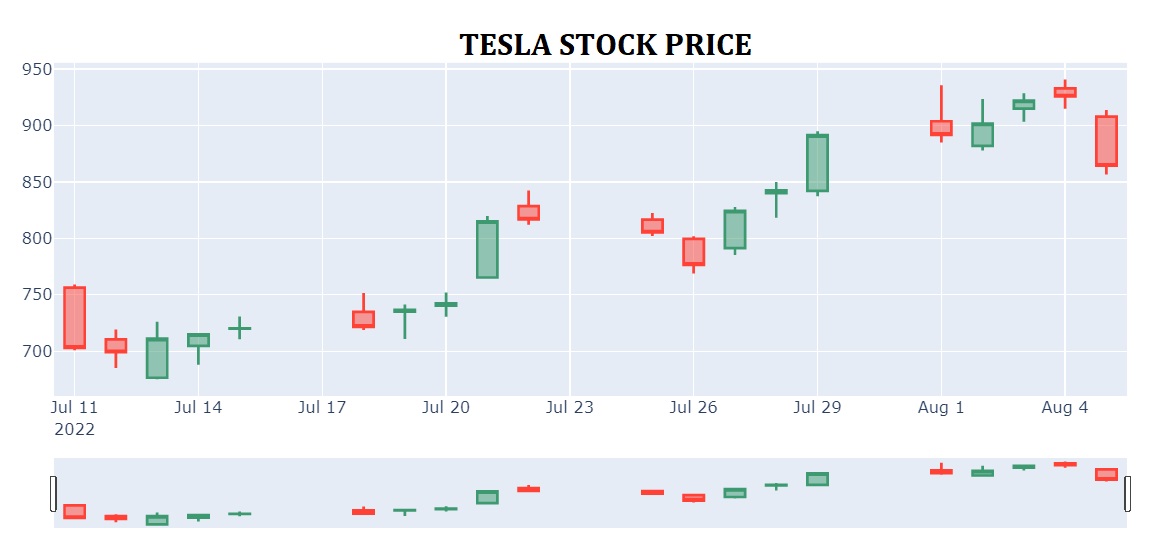}$\\
$\includegraphics[width = .45\textwidth]{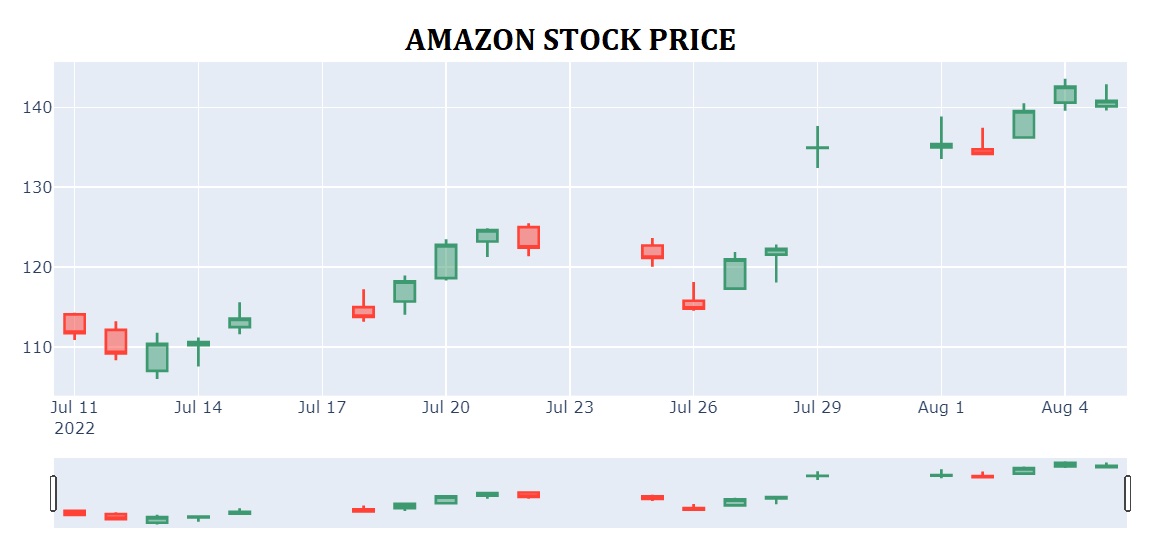}$ & $\includegraphics[width = .45\textwidth]{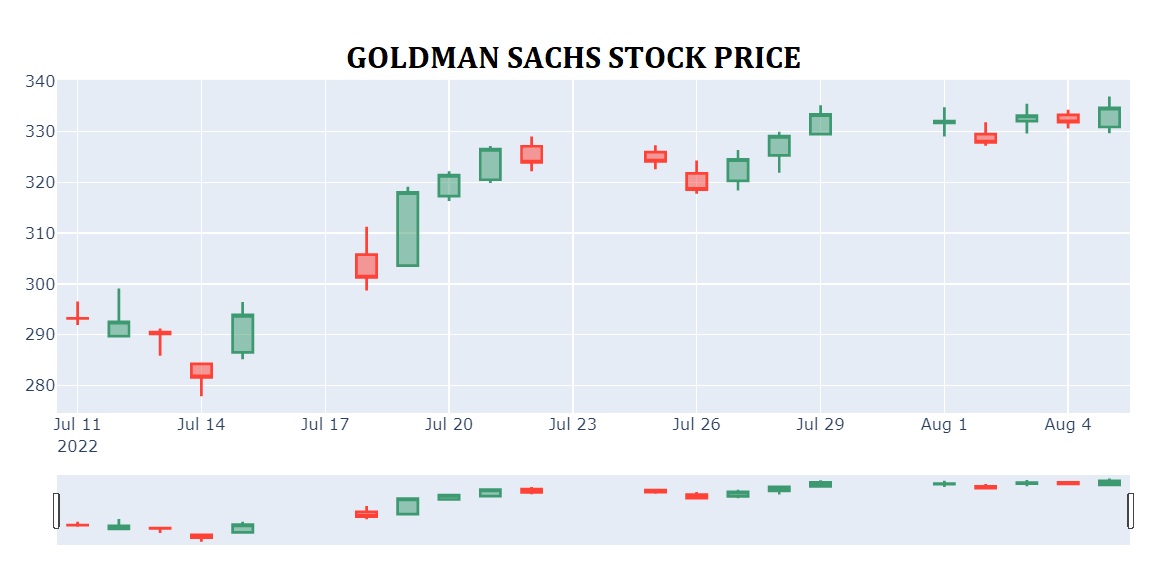}$\\ \\
\multicolumn{2}{c}{\textbf{Figure 1}. Candlestick charts for the four stocks over a period of 20 days}
\end{tabular*}
\]

\section{Findings and Analysis}\label{sec5}
\subsection{Sentiment analysis}

The evaluation of the sentiment expressed by each news piece is a critical component of our methodology for measuring how investors receive ESG information for investing decisions. As previously indicated, sentiment gathered from news articles should offer us the best proxy of a company's ESG achievements or deficiencies that are neither too skewed in favour of the firm's point of view nor heavily influenced by social media issues.

We use a dictionary technique to determine the sentiment conveyed by each unique news article. This approach uses pre-defined lists of terms that carry positive and negative meanings, respectively, to identify the sentiment indicated by a text. A popular approach is to rely on machine learning techniques. The advantage of computing sentiment using a machine learning technique is that one does not need to generate a positive and negative word list, but instead manually classifies a subsample of the data based on whether a certain text conveys positive or negative sentiment. The algorithm then learns from this training data to determine the most likely categorization of new articles by looking for patterns in the training data that are unique to either positive or negative sentiment. This advantage can also be viewed as a disadvantage of machine learning approaches because the quality of the sentiment index is heavily dependent on whether or not the training data contains a diverse set of features, i.e. words and linguistic structures that imply strong positive or negative sentiment (Siering, 2012). To avoid these limitations, we opt for a simple dictionary strategy. The pre-trained ProsusAI FinBERT model has been loaded to carry out the same. FinBERT is a pre-trained natural language processing (NLP) model that analyses the sentiment of financial documents. It is created by fine-tuning the BERT language model for financial sentiment categorization by further training it in the finance domain using a sizeable financial corpus. FinBERT is based on Hugging Face's `PyTorch pretrained bert' library and their BERT implementation for sequence classification problems. When applied to finance-related materials, their lexicon outperforms commonly used English language dictionaries, such as the General Inquirer's Harvard-IV-4 categorization dictionary. Calculating the sentiment index in this way leads to a variable that is defined on the interval $[-1, 1]$.

\subsection{Pre-trained ProsusAI/FinBERT}
FinBERT is an open-source pre-trained Natural Language Processing (NLP) model that has been trained particularly on financial data and surpasses practically all other NLP techniques for financial sentiment analysis.
Sentence context awareness is enabled through BERT. BERT is an acronym that stands for Bidirectional Encoder Representation from Transformer. It is one of Google's most popular state-of-the-art text embedding models. BERT has sparked a revolution in the realm of NLP by outperforming previous methods on numerous NLP tasks such as question answering, text generation, sentence classification, and many more.

FinBERT is a BERT-based language model. It improves the BERT model for financial data. The supplementary training corpus consists of 1.8 million Reuters news stories and the Financial PhraseBank. According to this paper [reference 1], the major sentiment analysis dataset used is Financial PhraseBank, which consists of 4845 English lines chosen at random from the LexisNexis database of financial news. Sixteen experts then analyzed these lines with finance and business expertise.

\subsection{Sentiment ‘Composite Score’ calculation and implications}
The Prosus/AI FinBERT dataset is only the first part of our analysis. After computation of the sentiment as `positive, negative or neutral' and the corresponding sentiment score through the pre-trained dataset, we have assigned weights. The weights allocation is as follows:
\begin{itemize}
\item Positive sentiment $= +1$
\item Neutral Sentiment $= 0$
\item Negative Sentiment $= (-1)$
\end{itemize}
After these weights have been assigned, the sentiment score initially computed is multiplied to create the `composite score' value for every news article obtained. After the composite score for each article has been calculated, the aggregate value is calculated based on segregation for each company. Figures 2, 3 and 4 present a sequential working of the problem, where a sentiment score is assigned to every ticker, then it is multiplied by the appropriate weight to get the composite score, which is then presented as a sentiment aggregate composite score which is calculated for each of the four stocks prior to standardization. The sentiment index has consistently favourable mean values. Given that this variable is constrained, all but one stock's sentiment appears to be negative. We can thus observe that for Amazon and Goldman Sachs, a strong positive aggregate sentiment is seen. For Tesla, a moderate positive aggregate sentiment is seen. However, for HSBC Bank, a strong negative sentiment is seen.

\begin{center}
\includegraphics[width = .8\textwidth]{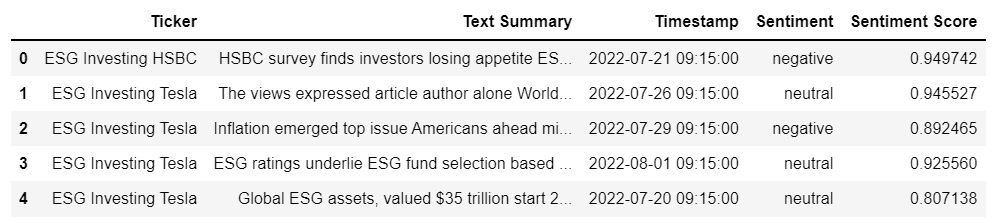}\\ 
\vspace{.2cm}
\textbf{Figure 2}. Sentiment analysis with sentiment score
\end{center}

\begin{center}
\includegraphics[width = .8\textwidth]{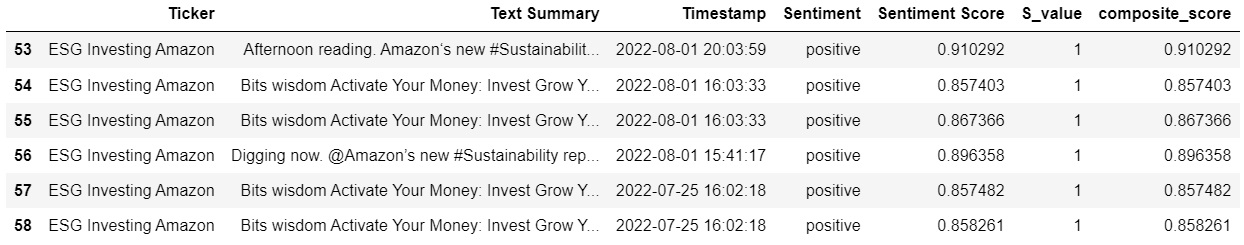}\\ 
\vspace{.2cm}
\textbf{Figure 3}. Providing a composite score
\end{center}

\begin{center}
\includegraphics[width = .6\textwidth]{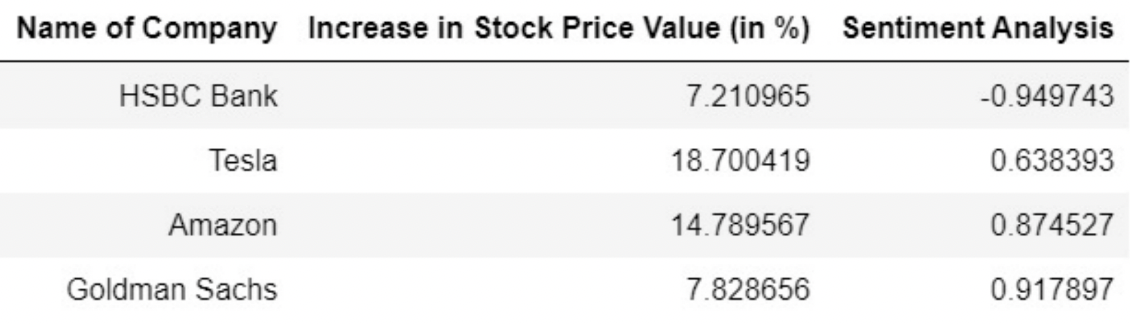}\\ 
\vspace{.2cm}
\textbf{Figure 4}. Relationship between stock prices and sentiment of ESG news
\end{center}

\subsection{Economic implications}
All factors taken into consideration, we can observe companies being `ESG-averse', `ESG-neutral', and `ESG-affine'. After a thorough examination, we can see that those organizations that have a high positive aggregate sentiment score can be classified as `ESG-affine'. In this case, Goldman Sachs, followed by Amazon, followed by Tesla, represents a decreasing order of `ESG-affine' organizations. Furthermore, we can observe that `ESG-averse' companies have a high negative aggregate sentiment score, which, in this case, is possessed by HSBC Bank.

\subsection{Limitations}
It is vital to acknowledge the drawbacks that exist, to make our research study a more comprehensive one. The primary drawback is the sample size of legitimate sources of news articles and Twitter data to compute the ESG sentiment was extremely low. Assessing the entire ESG sentiment of a company through just a few news articles and tweets (in some cases, lesser than 10) can lead to somewhat skewed conclusions. In future studies, one can consider a much larger sample size of social media networking data to calculate a more robust measure of the ESG sentiment score.

\section{Discussion}\label{sec6}
With an ever-increasing public and regulatory demand for publicly traded companies to adhere to socially responsible business practices and environmentally sustainable modes of operation, it appears that understanding how companies' efforts in the domains of environmental, social, and governance are perceived by investors and affect corporate performance is critical. Using twenty days of ESG-related news stories, we run machine learning models to see how the four equities (HSBC Bank, Tesla, Amazon, and Goldman Sachs) react to stock-specific ESG information. To assess the substance of each ESG-related news article, we extract its sentiment using the dictionary approach described by ProsusAI/FinBERT and then compute a polarity ESG-sentiment index for each trading day of each sample stock.

We can conclude from our research study that for the sample of ESG-related news articles for the Automobile, Online Retail, and Finance sectors, the stock prices have been impacted in a positive correlation by the ESG news published about the company, whereas the Banking sector is inversely correlated to the ESG news available on social media.

Our findings provide significant implications for businesses and investors alike, as well as fresh avenues for future research. First and foremost, businesses should pay special attention to their ESG initiatives. While our findings show that ESG activities should not be used carelessly to greenwash a company's public image and frequently fail to distract from financial under-performance, they do appear to have the potential to mitigate financial losses if investors' attitudes toward them are generally positive. Thus, our findings moderate the existing competing perspectives on the worth of ESG activities in academia and industry by demonstrating that ESG initiatives are context-dependent rather than unequivocally damaging or valuable. As a result, it would be beneficial to investigate the relationship between ESG initiatives and stock market returns in various scenarios. For example, the dramatic disparity in financial performance that we discover between the ESG-averse and ESG-affine groups of equities is something that both practitioners and researchers would be interested in. While data availability constraints prevented us from analyzing a broader sample of equities, access to larger data sets of news articles could provide additional insights into whether the trends we discovered also apply to S\&P 500 stocks. Larger samples may then enable the comparison or identification of industry clusters, as well as deeper correlations between market capitalization and ESG effects. While we purposefully ignored social media data, such information could be useful if the focus were shifted to the impact of ESG-related investor sentiment on returns or volatility. Finally, our findings should not be used to discourage corporations from engaging in socially responsible conduct but rather to highlight the topic's high degree of complexity, which necessitates a collaborative approach by both enterprises and investors to assess the genuine worth of ESG activities.

\bibliography{refer}

\end{document}